\documentclass[aps,prb,notitlepage,twocolumn]{revtex4-1}
\usepackage{graphicx}
\usepackage[colorlinks=true, allcolors=magenta]{hyperref}
\usepackage{xcolor}
\usepackage{amsmath,amssymb,bm}
\usepackage[utf8]{inputenc}

\usepackage{physics}
\usepackage{mathrsfs}
\usepackage{comment}
\usepackage{dcolumn}
\usepackage{mathptmx}
\usepackage{nccmath}

\renewcommand{\vr}{\mathbf{r}}
\newcommand{\vk}{\mathbf{k}}
\newcommand{\vv}{\mathbf{v}}
\newcommand{\vq}{\mathbf{q}}
\renewcommand{\d}{\partial}
\renewcommand{\to}{\widetilde{\omega}}

\renewcommand{\(}{\begin{equation}\begin{aligned}}
\renewcommand{\)}{\end{aligned}\end{equation}}

\begin{document}

\title{A generalized model of magnon kinetics and subgap magnetic noise}
\author{Haocheng Fang}
\author{Shu Zhang}
\email{suzy@physics.ucla.edu}
\author{Yaroslav Tserkovnyak}
\affiliation{Department of Physics and Astronomy, University of California, Los Angeles, California 90095, USA}
\date{\today}

\begin{abstract}
   Magnetic noise spectroscopy provides a noninvasive probe of spin dynamics in magnetic materials. We consider two-dimensional magnetically ordered insulators with magnon excitations, especially those supporting long-distance magnon transport, where nitrogen-vacancy (NV) centers enable the access to (nearly) ballistic transport regime of magnons. We develop a generalized theory to describe the magnon transport across a wide range of length scales. The longitudinal dynamic spin susceptibility is derived from the Boltzmann equation and extended to a Lindhard form, which is modified by both the spin-conserving magnon collisions and spin relaxation. Our result is consistent with the diffusive (ballistic) model for the length scale much larger (smaller) than the magnon mean free path, and provides a description for the intermediate regime. We also give a prediction for the NV transition rate in different magnon transport regimes.
\end{abstract}

\maketitle

\section{Introduction}

Magnons are quanta of spin-wave excitations in magnetically ordered systems. 
The study of magnon transport in magnetic insulators is both fundamentally and technologically motivated,~\cite{chumak2015review}
as the absence of electric currents may free the system from excess heating and noise.
Though conventional studies of magnon transport based on charge-spin conversion via metal-magnet contact~\cite{SSLZhang2012,jedema2001,cornelissen2015} are inspiring for an electrical control of magnetic devices, it is also desirable to be able to directly probe the intrinsic magnetic transport properties without the complexity of interfacial effects.
A noninvasive probe can be achieved by detecting the fluctuations of the magnetic field emitted into the environment, which result from the presence of magnetic excitations. Recently, the development of single-qubit relaxometry using NV centers~\cite{NV2008,Yacoby2018} offers the advantages of a high frequency resolution in the GHz scale and a tunable  NV-sample distance in a wide range from nanometers to micrometers. The latter, in particular, enables the probe of transport on different length scales and scattering regimes.~\cite{Demler2017Graphene,Rodriguez-2018,Chatterjee2019}

Generally speaking, the NV relaxation rate is highly sensitive to the magnetic noise generated by spin fluctuations, which is related to the spin response functions by the fluctuation-dissipation theorem.
Considering the magnetic noise at the position of the NV sensor, the contribution from different length scales in the material are weighted by a form factor, which is determined by the demagnetization kernel, i.e., the spatial distribution of the stray field generated by a magnetic moment. The weight peaks at the length scale comparable to the distance $d$ from the NV to the material surface.~\cite{Yacoby2018}

For a magnon gas, a key relevant length scale is the magnon mean free path $l$. From the perspective of the NV at $d \ll l$, the magnons propagate freely without encountering collisions. In the opposite limit, $d \gg l$, the averaged effects of random collisions give rise to magnon diffusion. 
Magnetic insulators with long-distance magnon transport have recently attracted much attention, where the magnon diffusion length can reach a few micrometers.~\cite{Myers2015,cornelissen2015,lebrun2018,Wang2020} In clean systems, magnon mean free paths are expected to be long as well, which can be comparable with or exceed $d$. 
Therefore, NV measurement of these materials is promising to approach and study the (nearly) ballistic and perhaps hydrodynamic~\cite{rodriguez2018hydrodynamic} regimes of magnon transport.

\begin{figure}[b]
    \centering
    \includegraphics[width = \linewidth]{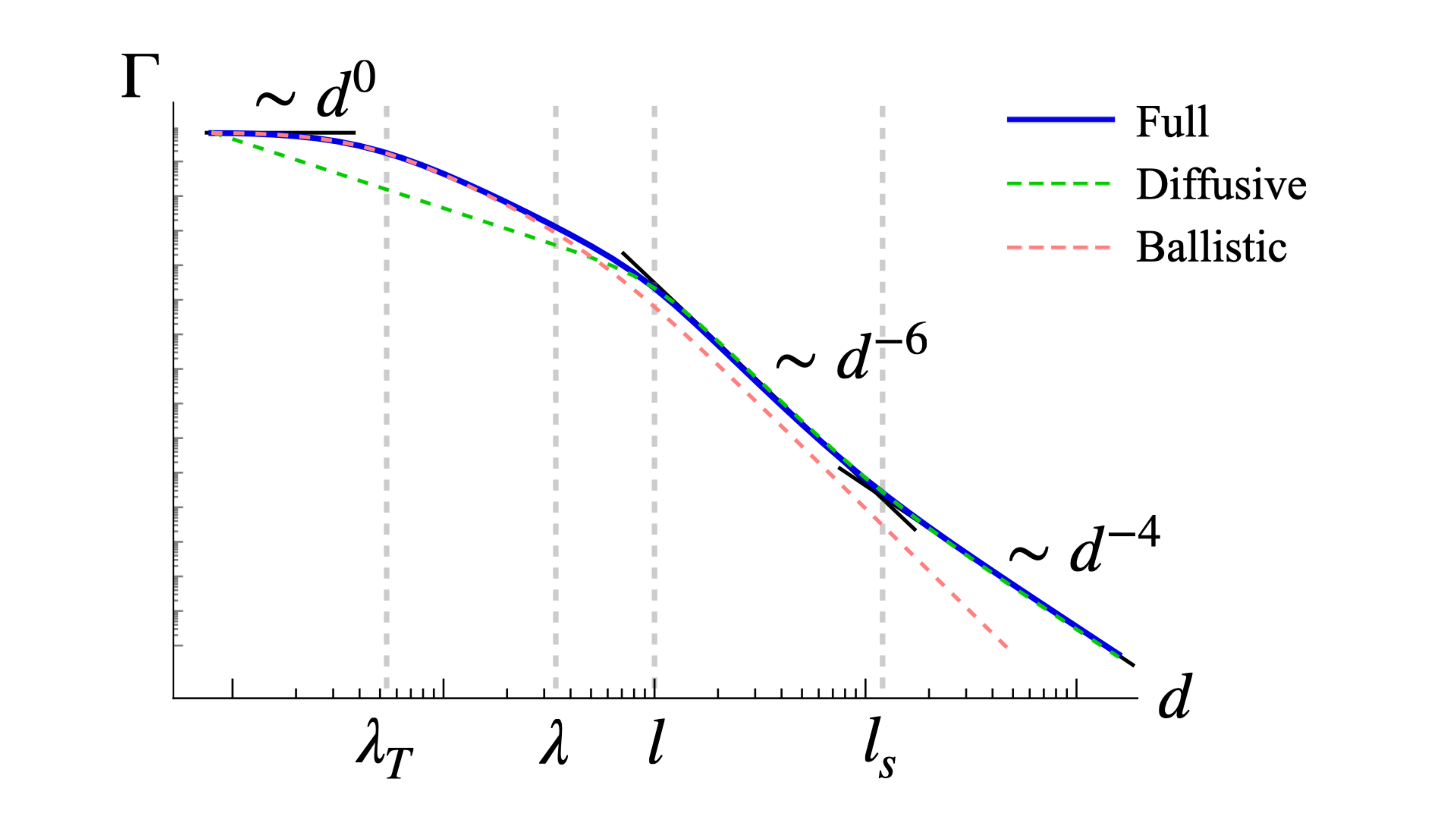}    
    \caption{Log-log plot of the NV transition rate scaling with the NV distance $d$ across different magnon transport regimes. The length scales in the plot are the magnon diffusion length $l_s$, the magnon mean free path $l$, the characteristic wavelength $\lambda$ of the order of the magnetic healing length, and the wavelength $\lambda_T$ of the thermal magnons with energy $k_B T$. Effective parameters in the diffusive model are derived from the full model in the diffusive limit. See main text and Appendix A for details.
    Here, the frequency $\omega$ is set to be the zero-field NV electron spin resonance frequency $\omega/2\pi = 2.87$~GHz. Plots with two other choices of frequency are given in Appendix B.
    }
    \label{fig1}
\end{figure}

In this work, we develop a generalized theory for free magnon transport, which is applicable in the ballistic, diffusive and intermediate regimes. In the Boltzmann equation for the magnon wave packet, we include both magnon-number conserving and nonconserving scattering processes under relaxation-time approximation. The resulting dynamic spin susceptibility in a modified Lindhard form reduces to the known results in ballistic and diffusive limits. By comparing with the phenomenological spin diffusion result, important transport properties such as the spin diffusion constant and the spin conductivity can be estimated in terms of microscopic parameters in the magnon dispersion.
We focus on the scenario where the energy scale of the NV resonance frequency $\hbar \omega$ $\ll$ the magnon gap $\Delta$ $\ll$ $k_B T$, where $k_B$ is the Boltzmann constant and $T$ is the temperature. Our analysis shows that the transport in two dimensions is mainly contributed by low-energy (infrared) magnons, rather than the thermal (ultraviolet) magnons with energy $k_B T$,  consistent with experimental reports on the importance of subthermal magnons in spin transport.~\cite{Boona2014,Jin2015}
Accordingly, a characteristic magnon wavelength $\lambda$ of the order of the magnetic healing length $\sqrt{A/\Delta}$, where $A$ is the exchange stiffness,  arises as another length scale, in addition to the magnon mean free path $l$ and diffusion length $l_s = \sqrt{D \tau}$, where $D$ and $\tau$ are the the spin diffusion constant and the spin relaxation time, respectively, in a spin diffusion description.  Our main results are illustrated in Fig.~\ref{fig1}, where the distance dependence  of the NV relaxation rate is shown over a wide range. Our theory provides a general prediction for the subgap NV probe of magnon transport in magnetic insulators.

\begin{figure}[t]
    \centering
    \includegraphics[width = \linewidth]{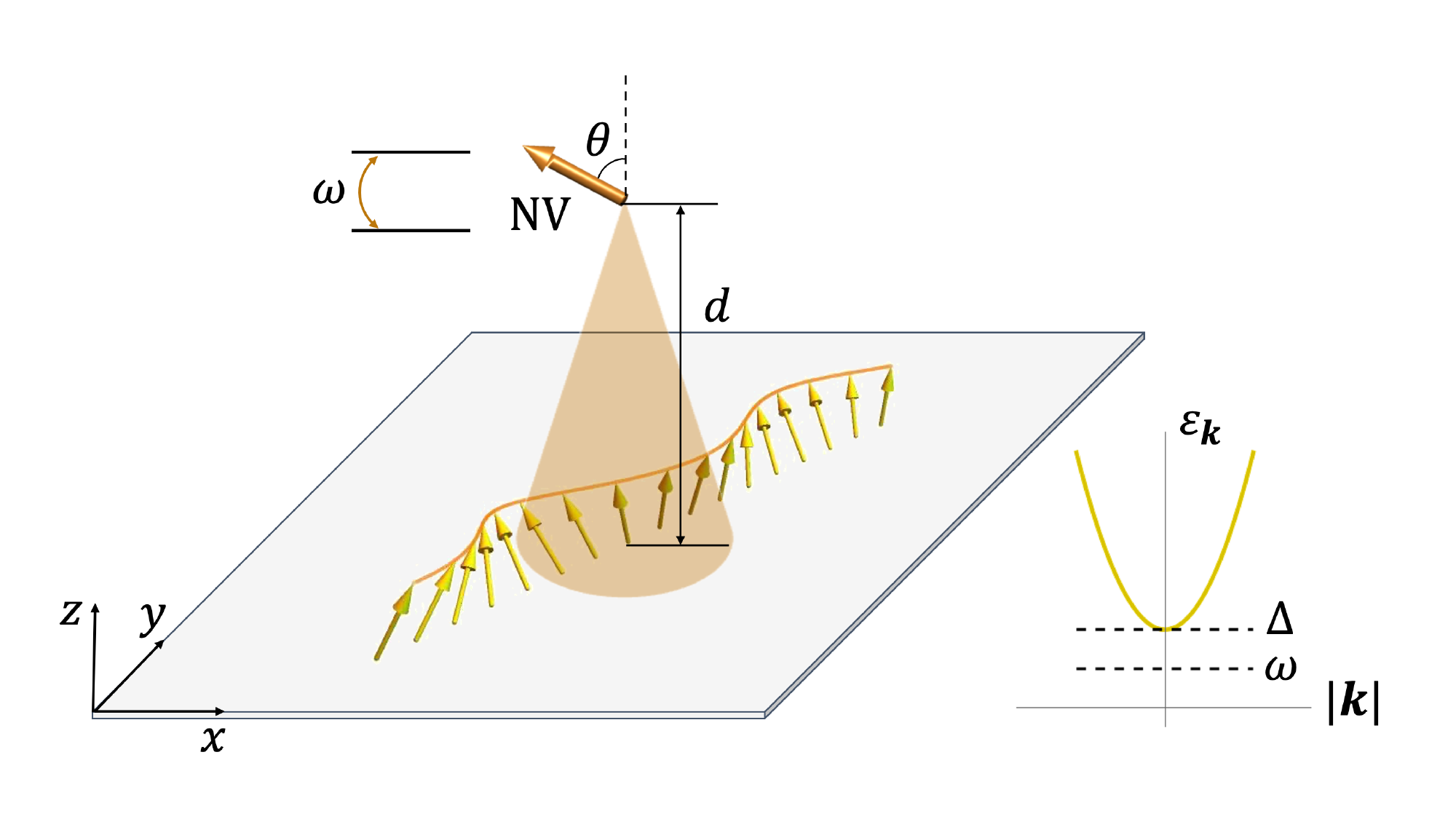}
    \caption{Schematic for subgap NV measurement of a two-dimensional magnetic insulator with magnon excitations. The NV is placed distance $d$ above the magnetic thin film with orientation angle $\theta$ relative to the $z$ axis. The NV resonance frequency $\omega$ is smaller than the magnon gap $\Delta$. Inset: the quadratic dispersion of magnons.}
    \label{fig2}
\end{figure}

\section{Model}

Let us consider the setup as depicted in Fig.~\ref{fig2}. An NV center is placed at distance $d$ from a two-dimensional magnetic thin film, which is collinearly ordered with its order parameter, magnetization for example, along the easy axis $\hat{\mathbf{z}}$. 
Both transverse and longitudinal spin fluctuations generate magnetic noise. For the former, the NV transition can be accompanied by the creation or annihilation of one magnon at the NV resonance frequency, and the magnon spectrum can thus be mapped out by tuning the NV frequency.~\cite{vandersar2015,wolf2016} 
The latter, on the other hand, is induced by multimagnon processes,~\cite{Hammel2020,Du2020} for example, one magnon scattering into another with an energy gain or loss equal to the NV frequency.
For many magnetic materials, the magnon gap can be much larger than the NV resonance frequency (typically a few GHz) as in our case. The transverse coupling is suppressed and the NV measurement is dominated by the longitudinal spin dynamics, which reflects the population distribution of the magnon gas, magnon collisions and spin relaxation effects, etc.
The NV transition rate $1/T_1$ is given by~\cite{vandersar2015,Flebus2018}
\(\label{eq:transition-rate}
    \Gamma (\omega,d,\theta) 
    = (\gamma \widetilde{\gamma})^2 f(\theta) 
    \frac{k_B T}{\hbar \omega} \int dq\, q^3 e^{-2qd} \chi'' (q,\omega),
\)
where $\chi'' (q,\omega)$ is the longitudinal dynamic spin susceptibility at frequency $\omega$ and wavenumber $q$, assuming axial symmetry in the spin space. To obtain this form, the fluctuation dissipation theorem
$C_{zz} (q, \omega) 
= (2k_B T/ \hbar \omega) \chi'' (q,\omega)$ has been invoked in its classical form for high temperatures $k_B T\gg \hbar \omega$, where $C_{zz} (q, \omega)$ is the correlation function of the longitudinal spin component $s_z$.
Here, $\gamma$ and $\widetilde{\gamma}$ are the gyromagnetic ratio of the magnetic film and the NV spin, respectively, and $f(\theta) =  (5-\cos 2\theta)\pi/2$ is an overall factor determined by the orientation angle $\theta$ of the NV spin. This geometric factor is modified when the magnetic order is aligned differently.~\cite{Wang2020}
The filter function $q^3 e^{-2qd}$, as a result of the demagnetization kernel discussed above, shows that the NV relaxation is dominated by contribution from spin dynamics at length scale $d$.
Therefore, the noise is mainly characterized at wavenumber $q \sim 1/d$, while magnons from all wavevectors $\vk$ in general contribute, as we shall see later in the treatment of the Boltzmann equation. 
We now turn to the study of the dynamic spin susceptibility in different transport regimes of a magnon gas.
For simplicity, we focus on the exchange-dominated isotropic magnon transport, neglecting Hall effects. 

\section{Dynamic susceptibility}

\subsection{Diffusive regime}

Suppose the temperature is uniform and quickly relaxing. In the diffusive regime ($d \gg l$), the natural dynamics of the deviation of the magnon density $\delta n$ from the thermal equilibrium can be described by the diffusion equation
\(
\partial_t \delta n + \bm{\nabla} \cdot \mathbf{j}_s = - \frac{\chi_0}{\tau} \mu,
\label{eq:diffusion}
\)
where $\chi_0$ is the uniform static longitudinal spin susceptibility, $\tau$ is the spin relaxation time, and $\mathbf{j}_s = -\sigma \bm{\nabla} \mu $ is the spin current. As in Fick's law, we have the spin conductivity $\sigma$ and the spin chemical potential $\mu =  \delta n/\chi_0 - h$,
where $h$ is a force thermodynamically conjugate to $\delta n$, which may arise from an external magnetic field. As the magnon density $\delta n$ relaxes towards $\chi_0 h$,
we have
\(
\left(\partial_t + \frac{1}{\tau}\right) \delta n - D \nabla^2 \delta n  = \chi_0 \left( - D \nabla^2  + \frac{1}{\tau}\right)h, 
\label{eq:diffusion-h}
\)
where $D = \sigma/\chi_0$ is the spin diffusion constant.

Taking a single Fourier component of the external perturbation
$h(\vr,t) = h (\vq, \omega) e^{i \vq \cdot \vr - i \omega t}$, in the linear response, 
$\delta n (\vr, t) = \delta n (\vq, \omega) e^{i \vq \cdot \vr - i \omega t}$ follows the same modulation. The dynamic spin susceptibility is given by $\delta n (\vq, \omega) = \chi_D (\vq, \omega) h (\vq, \omega)$, where
\(  \label{eq:diffusive-susceptibility}
    \chi_D
    =  \frac{\chi_0 (D q^2 + 1/\tau)}{-i\omega  + Dq^2 + 1/\tau}.
\)
When the spin relaxation effect dominates the magnon kinetics, $q l_s \ll 1$, where $l_s = \sqrt{D \tau}$ is the spin diffusion length,  the expression simplifies, $\chi_D \rightarrow \chi_0/(-i \omega \tau +1)$. In the fast relaxation limit $\tau \rightarrow 0$, $\chi_D \rightarrow \chi_0$, the system always stays in equilibrium with $\mu = 0$.

\subsection{Ballistic regime}

In the opposite limit ($d \ll l$), we consider the ballistic transport of magnons, following the Lindhard treatment of the following Hamiltonian of a magnon gas:
\(
    H &=H_0 +H_1 \\
    &= \sum_\vk \epsilon_\vk a_\vk^\dagger a_\vk
    - \int d^2  \vr \, h(\vr, t) a^\dagger(\vr) a(\vr),
    \label{eq:Hamiltonian}
\)
where $a_\vk^\dagger$ and $a_\vk$ are the (bosonic) magnon operators, 
$a(\vr) 
= (1/\sqrt{V}) \sum_\vk a_\vk e^{i \vk \cdot \vr}$,
$\epsilon_\vk$ is the magnon dispersion, and $h(\vr, t)$ is a perturbing potential in favor of a nonvanishing local magnon density.

We again focus on one Fourier component $h(\vq,\omega)$ of the perturbation, and define the density operator 
$\rho_\vq = \int d^2 \vr \, a^\dagger(\vr) a(\vr) 
e^{-i \vq \cdot \vr} = (1/V) \sum_\vk \rho_{\vk,\vq}$, where
$\rho_{\vk, \vq} = a_{\vk-\vq/2}^\dagger a_{\vk+\vq/2}$. The perturbation term thus reads $H_1 = (1/V) \sum_\vk \rho_{\vk,-\vq} h(\vq, \omega) e^{-i\omega t}$. The time evolution of $\rho_{\vk,\vq}$ follows the Heisenberg equation
$ i\hbar(d/dt) \rho_{\vk,\vq} = [\rho_{\vk,\vq}, H]$, which yields
\(
    (\hbar \omega + i 0^+) \langle \rho_{\vk,\vq} \rangle
    &= (\epsilon_{\vk+\vq/2} - \epsilon_{\vk-\vq/2}) 
    \langle \rho_{\vk,\vq} \rangle \\
    & + (f^0_{\vk+\vq/2} - f^0_{\vk-\vq/2} ) h(\vq, \omega). 
\)
Here, $f^0_\vk = \langle a_\vk^\dagger a_\vk \rangle = 1 \big/\left( e^{\beta \epsilon_\vk } -1 \right)$ is the Bose-Einstein distribution of the magnon gas in equilibrium, $\beta = 1/k_B T$, and $i0^+$ is a regularization to suppress ringing. Since $\delta n(\vq, \omega) = (1/V) \sum_\vk \langle \rho_{\vk,\vq} \rangle$, we arrive at the Lindhard function for the dynamic susceptibility of magnons:
\( \label{eq:ballistic-susceptibility}
    \chi_B 
    = - \int \frac{d^2 \vk}{ 4 \pi^2}
    \frac{f^0_{\vk+\vq/2} - f^0_{\vk-\vq/2} }
    {\epsilon_{\vk+\vq/2} - \epsilon_{\vk-\vq/2} - \hbar \omega - i 0^+},
\)
where we have replaced the sum over wavevectors by the integration.

\subsection{General form}

Studies by Kliewer and Fuchs~\cite{Kliewer1969} and  Mermin~\cite{Mermin1969} have shown that the Lindhard function of an electron gas can be generalized in the presence of electron collisions under the relaxation-time approximation.
Here, we derive a general form of the dynamic spin susceptibility of the magnon gas, taking into account both magnon collisions and spin relaxation. The final result, extended to a Lindhard from, is consistent with the forms in both diffusive and ballistic limits, as given above.

First, in the long-wavelength limit, we describe the magnons as semiclassical wave packets with a distribution
$f_\vk (\vr,t)$ in the phase space, the dynamics of which is captured by the Boltzmann equation:
\(\label{eq:Boltzmann-equation}
   \left[
   \frac{\d}{\d t} 
   + \mathbf{v}_\vk \cdot \frac{\d}{\d \vr} 
   + \frac{\d h(\vr,t)}{\d \vr} \!\cdot\! \frac{\d}{\d \hbar \vk} 
   \right] 
   f_\vk (\vr,t)
   = \left( \! \frac{\d f_\vk }{\d t} \! \right)_{\!\!c}
   +\left( \! \frac{\d f_\vk }{\d t} \! \right)_{\!\!r} . 
   \\[2pt]
\)
Here, $\mathbf{v}_\vk = \d \epsilon_\vk / \d \hbar \vk$ is the group velocity of the magnon wave packet and $h(\vr,t)$ is a slowly varying potential [consistent with those in Eq.~(\ref{eq:diffusion-h}) and Eq.~(\ref{eq:Hamiltonian})] driving the motion of the wave packet. 
The two scattering terms on the right-hand side respectively characterize collisional processes that conserve the magnon number and relaxational processes that do not. The former refers to the spin-conserving exchange-dominated scattering between magnons, including Umklapp processes, spin-conserving magnon-phonon scattering, and scattering by nonmagnetic disorder. These processes can be elastic and inelastic, which locally redistribute the momentum and energy of magnons while preserving the magnon number. The latter corresponds to magnon decay as a result of those magnon-phonon and magnon-magnon scattering processes that create or annihilate magnons, arising from spin-orbit couplings and magnetic dipole-dipole interactions. 
When spin-conserving scattering processes are dominant, a magnon chemical potential can be introduced in the magnon distribution, which has been useful for the description of the spin Seebeck effect,~\cite{cornelissen2015,Cornelissen2016} magnon Bose-Einstein condensation,~\cite{demokritov2006bose,giamarchi2008bose,Bender2012} etc. 
As we shall see, the local magnon chemical potential is crucial in our treatment, which enables a self-consistent solution to the Boltzmann equation under the relaxation-time approximation. The collective energy dynamics is neglected in this work by taking the temperature as a uniform constant. We primarily consider the magnon energy relaxation into a phonon bath, as phonons can have a relatively high heat conductivity around the room temperature. 

Assuming the relaxation-time approximation is still valid, we introduce a collision time $\tau_c$ and a relaxation time $\tau_r$ for the spin conserving and non-conserving scattering processes, respectively.
Both $\tau_c$ and $\tau_r$ are considered to be energy dependent, though we drop the explicit $\vk$ dependence in notation.
The relaxation term can be treated straightforwardly, 
\( \label{eq:relaxation}
    \left( \! \frac{\d f_\vk }{\d t} \! \right)_{\!\!r}
    = -\frac{f_\vk - \widetilde{f}^0_\vk}{\tau_r},
\)
which relaxes the magnon distribution towards 
$\widetilde{f}^0_\vk = 1\big/\left\{e^{\beta [\epsilon_\vk - h (\vr, t)]} - 1\right\} \approx f^0_\vk - (\partial f^0_\vk /\partial \epsilon_\vk) h(\vr,t)$ according to the external perturbation.
Following the work by Kragler and Thomas on the modified Lindhard function for electrons,~\cite{Kragler1980} 
let us introduce a local magnon chemical potential $ \mu(\vr,t)$ to handle the collision term, which relaxes the system towards a (fictitious) local equilibrium distribution
$g_\vk (\vr, t) 
= 1\big/\left\{e^{\beta \left[\epsilon_\vk - \mu(\vr,t)\right]} -1 \right\}$,
\(  \label{eq:collision}
    \left( \! \frac{\d f_\vk }{\d t} \! \right)_{\!\!c}
    = -\frac{f_\vk - g_\vk}{\tau_c}.
\)
This describes the thermalization of magnons among different energy modes by redistributing their momenta, and at the same time, enables us to implement the conservation of the local magnon number $n(\vr,t) = \int (d^2 \vk /4\pi^2) f_\vk$: The integration of Eq.~(\ref{eq:collision}) over all wavevectors must vanish.
By expanding
$f_\vk (\vr, t) 
= f^0_\vk  +  \delta f_\vk (\vr, t)$, 
and
$g_\vk (\vr, t) 
= f^0_\vk  +  \delta g_\vk (\vr, t) 
\approx f^0_\vk  -(\d f^0_\vk / \d \epsilon_\vk)  \mu(\vr,t)$ to linear order, the local chemical potential is determined self-consistently,
\( \label{eq:chemical-potential}
\mu(\vr,t) 
= -\frac{
\int (d^2 \vk /4\pi^2) \delta f_\vk /\tau_c}
{\int (d^2 \vk /4\pi^2) 
(\d f^0_\vk / \d \epsilon_\vk) /\tau_c}.\)
As a reminder, $\tau_c$ is generally $\vk$ dependent and cannot be taken out of the integral. 

Applying the relaxation-time approximations~(\ref{eq:relaxation}, \ref{eq:collision}) to the Boltzmann equation~(\ref{eq:Boltzmann-equation}) leads to its following form in the momentum space
\begin{align}
    & \left( 
    -i\omega + i \vq \cdot \mathbf{v}_\vk + \frac{1}{\tau_c} + \frac{1}{\tau_r}
    \right)
    \delta f_\vk (\vq, \omega) 
     + \frac{1}{\tau_c}  \left(\frac{\partial f_\vk^0}{\partial \epsilon_\vk} \right) \mu (\vq,\omega) \nonumber\\
    & \qquad \qquad
    = 
    - \left( i \vq \cdot \mathbf{v}_\vk +\frac{1}{\tau_r} \right)
    \left( \! \frac{\d f^0_\vk}{\d \epsilon_\vk} \! \right) 
    h (\vq, \omega).  \label{eq:Boltzmann-k}
\end{align}
The equation above is first integrated to eliminate $\delta f$, according to the self-consistent condition (\ref{eq:chemical-potential}), to obtain the response of the local chemical potential to the external perturbation
$ \mu (\vq,\omega) 
= M(\vq,\omega) h(\vq,\omega)$, where
\begin{align}
    &M(\vq,\omega) = 
    {\displaystyle \int} \cfrac{d^2 \vk}{ 4 \pi^2} \cfrac{i \vq \cdot \mathbf{v}_\vk +1/\tau_r}{-i \omega + i \vq \cdot \mathbf{v}_\vk + 1/\tau_c +1/\tau_r} \cfrac{1}{\tau_c}
    \left(\cfrac{\partial f_\vk^0}{\partial \epsilon_\vk} \right) \nonumber\\
    & \times \left[ {\displaystyle \int} 
    \cfrac{d^2 \vk}{ 4 \pi^2} \cfrac{-i \omega + i \vq \cdot \mathbf{v}_\vk + 1/\tau_r}{-i \omega + i \vq \cdot \mathbf{v}_\vk + 1/\tau_c +1/\tau_r}\cfrac{1}{\tau_c} \left(\cfrac{\partial f_\vk^0}{\partial \epsilon_\vk} \right) \right]^{-1} \hspace{0pt}.\label{}
\end{align}
Reintegrating Eq.~(\ref{eq:Boltzmann-k}) with the response $\mu (\vq, \omega)$ substituted in yields the dynamic spin susceptibility
\(  \label{eq:boltzmann-susceptibility}
    &\chi
    = - \int \frac{d^2 \vk}{ 4 \pi^2} \frac{i \vq \cdot \mathbf{v}_\vk + 1/\tau_r + (1/\tau_c)  M(\vq,\omega)}{-i \omega + i \vq \cdot \mathbf{v}_\vk + 1/\tau_c +1/\tau_r}
      \left(\frac{\partial f_\vk^0}{\partial \epsilon_\vk} \right). \\[2pt]
\)

At the linear order in $\vq$, 
$\epsilon_{\vk+\vq/2} - \epsilon_{\vk-\vq/2} 
\approx \hbar \vq \cdot \mathbf{v}_\vk$, which is exact for a quadratic dispersion relation and is generally applicable for $q \ll k$, 
\( \label{eq:recover-lindhard}
\frac{\d f^0_\vk }{ \d \epsilon_\vk }
\approx \frac{f^0_{\vk+\vq/2} - f^0_{\vk-\vq/2}}{\epsilon_{\vk+\vq/2} - \epsilon_{\vk-\vq/2}}.
\)
One therefore recognizes the Lindhard form of the result (\ref{eq:boltzmann-susceptibility}) in the absence of scattering, which reduces to
\(
    \chi
     \rightarrow  \frac{ - L_0(\vq, \omega + i/\tau_c )}
    {1 - 
    [1/(1 - i\omega \tau_c )]
    [1 - L_0(\vq, \omega + i/\tau_c )/L_0(\vq, 0)]},
\)
for a constant $\tau_c$ and in the limit $1/\tau_r \rightarrow 0$,
where $L_0(\vq, \omega) = -\chi_B$ is the bare Lindhard function as given by Eq.~(\ref{eq:ballistic-susceptibility}). 
This is exactly the form of Mermin's result for the modified Lindhard dielectric function of an electron gas under relaxation-time approximation.~\cite{Mermin1969}
For magnons following the Bose-Einstein distribution function, the derivative $(\d f^0_\vk / \d \epsilon_\vk)$ grows towards the bottom of the magnon band, and is significant for all subthermal energies.
This is very different from the case with electrons, where the states around the Fermi surface dominate.
Therefore, it is necessary to account for the energy dependence of the relaxation time $\tau_c$ and $\tau_r$ inside the integrals. 

To extend our result eliminating the long-wavelength assumption, we restore the Lindhard form by applying Eq.~(\ref{eq:recover-lindhard}) and replacing $\vq \cdot \mathbf{v}_\vk$ by $(\epsilon_{\vk+\vq/2} - \epsilon_{\vk-\vq/2})/\hbar$. This leads to a full expression for the dynamic susceptibility applicable in all of the discussed regimes:
\( \label{eq:full-susceptibility}
    \chi_F  (\vq, \omega)
    = &- \widetilde{L}_0 (\vq,\omega)  + \widetilde{N}_{0,1} (\vq,\omega) \\
    &  + \cfrac{  \widetilde{N}_{1,0} (\vq,\omega)
    \left[ \widetilde{L}_1 (\vq,\omega) - \widetilde{N}_{1,1} (\vq,\omega)\right] }   {L_1(\vq,0) \!+\! \widetilde{N}_{2,0}(\vq,\omega)},
\)
where
\begin{align}
    L_m(\vq,\omega) 
    &=
    {\displaystyle \int} 
    \cfrac{d^2 \vk}{ 4 \pi^2} \left(\cfrac{ 1 }{\tau_c}\right)^{\!\!m}
    \cfrac{f^0_{\vk+\vq/2} - f^0_{\vk-\vq/2} }
    {\epsilon_{\vk+\vq/2}- \epsilon_{\vk-\vq/2} - \hbar\omega},
    \quad \text{and} \nonumber \\
    N_{m,n} (\vq,\omega) 
    &= {\displaystyle \int} 
    \cfrac{d^2 \vk}{ 4 \pi^2} 
    \left(\cfrac{ 1 }{\tau_c}\right)^{\!\!m}
    \left(\cfrac{ 1 }{\tau_r}\right)^{\!\!n} \nonumber\\
    & \hspace{-20pt} \times \cfrac{i\hbar \left(f^0_{\vk+\vq/2}- f^0_{\vk-\vq/2}\right)}
    { \left(\epsilon_{\vk+\vq/2} - \epsilon_{\vk-\vq/2} - \hbar \omega \right)\left(\epsilon_{\vk+\vq/2} - \epsilon_{\vk-\vq/2}\right)},
\end{align}
with $m, n = 0,1,2$. The tilded symbols $\widetilde{L}_m (\vq,\omega)$ and $\widetilde{N}_m (\vq,\omega)$  denote ${L}_m (\vq,\omega)$ and ${N}_m (\vq,\omega)$, respectively, with  $\omega$ replaced by $\to = \omega + i/\tau_c + i/\tau_r$ in their integrands. By doing so, we exactly recover the ballistic result~(\ref{eq:ballistic-susceptibility}) in the limit of $\tau_c, \tau_r \rightarrow \infty$.

It can also be shown that the full susceptibility~(\ref{eq:full-susceptibility}) reduces to the previous result in the diffusive limit~(\ref{eq:diffusive-susceptibility}),
from which we can extract the effective transport coefficients in terms of microscopic parameters. The details are shown in Appendix A.
As a simple but natural scenario, we assume the magnon mean free path $l$ is energy independent, giving a $\vk$ dependent collision time
$\tau_c = l/|\mathbf{v}_\vk|$.
The relaxation time is also energy dependent $\tau_r = \hbar / 2\alpha \epsilon_\vk$, where $\alpha$ is the Gilbert damping. 
For a quadratic dispersion relation $\epsilon_\vk = A \mathbf{k}^2 + \Delta$, 
we obtain the spin diffusion constant
$D \approx {\pi l \sqrt{A\Delta} / 2 \hbar}$, the spin conductivity
$\sigma  \approx { k_B T l / 8 \hbar \sqrt{A\Delta}}$, and the spin relaxation time
$\tau  \approx {\hbar / 2\alpha \Delta \ln\left(k_B T / \Delta\right)}$.
These parameters are used to plot the diffusive curve in Fig.~\ref{fig1}, giving excellent agreement with the general result at $d \gg l$.

It is worth remarking on the physical implications of these expressions. 
In general, spin transport can be quantitatively sensitive to the scattering regime. Here, the energy dependence of the collision time $\tau_c \sim 1/|\vk|$ accentuates low-energy magnons in two-dimensional magnon transport. As a result, the spin conductivity can be enhanced by a small magnon gap compared with the three-dimensional case, which may be relevant to the recent experimental observations.~\cite{wei2021}
From the two-dimensional spin diffusion constant $D$, an effective velocity of magnons contributing to the diffusive spin transport can be introduced $v = 2D/l = \pi\sqrt{A\Delta}/ \hbar$, which is much lower than the velocity of the thermal magnons with energy $k_B T \gg \Delta$.   Intuitively, due to the energy dependence of the Bose-Einstein distribution, a large population of low-energy magnons participate in the spin transport, even at the temperature regime that is of the same order of (but lower than) the magnetic ordering temperature.
The corresponding magnon wavelength $\lambda = 4\sqrt{A/\Delta}$, which is of the order of the magnetic healing length, emerges as another length scale in the spin transport. Although much larger than the thermal-magnon wavelength, it is still much smaller than the magnon mean free path, consistent with the semiclassical description and relaxation-time approximation. The collision time of these characteristic low-energy magnons can also be taken as an effective collision time constant in the system: $\tau_0 = \tau_c (|\vk| = 2\pi/\lambda) = l/v = \hbar l/\pi \sqrt{A \Delta}$, which determines the frequency range  $\omega \tau_0 \ll 1$ of the validity of the diffusive model.

\begin{figure}[b]
    \centering
    \includegraphics[width = \linewidth]{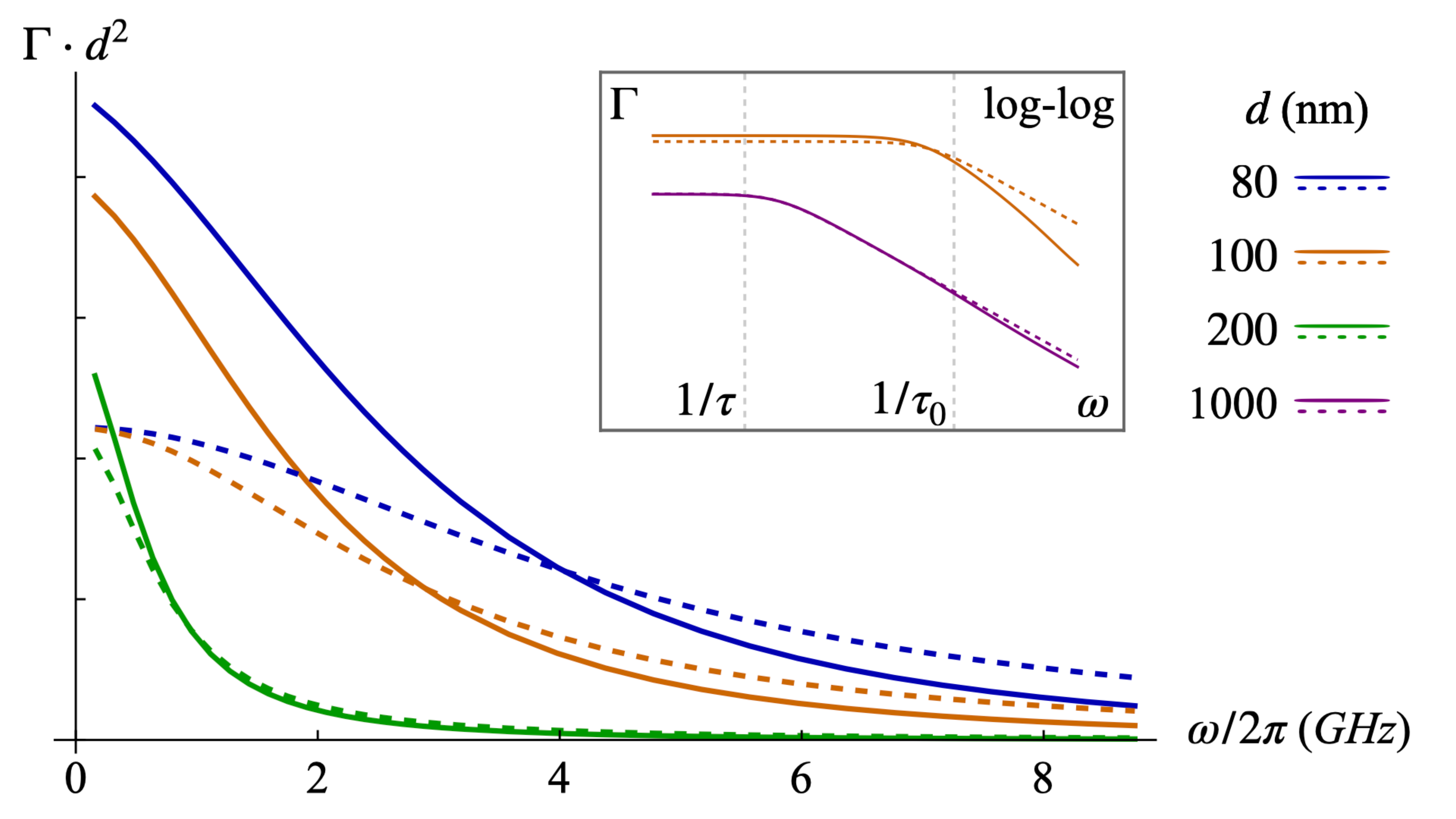}
    \caption{NV transition rate when tuning the NV resonance frequency, plotted on a linear scale, for distances $d = 80, 100  ~(=l)$, and $200$~nm. Solid (dashed) lines are calculated from the full (diffusive) model. Inset: log-log plot for $d = 100$ and $1000$~nm. The frequencies corresponding to the effective spin relaxation rate $1/\tau$ and effective magnon collision rate $1/\tau_0$ are shown. }
    \label{fig3}
\end{figure}

\section{Experimental prediction}

We are now ready to estimate the NV transition rate by substituting the results of the dynamic susceptibility into Eq.~(\ref{eq:transition-rate}).
In Fig.~\ref{fig1}, we plot the transition rate calculated from the full theory~(\ref{eq:full-susceptibility}) scaling with the NV distance $d$ across the ballistic, intermediate, and diffusive regimes. It agrees well with the diffusive result~(\ref{eq:diffusive-susceptibility}) for $d \gg l$ and with the ballistic result~(\ref{eq:ballistic-susceptibility}) for $d \ll l$, and provides a prediction for the intermediate regime, where both the diffusive and ballistic descriptions break down.

For the numerical plots, we have chosen the spin stiffness $A = 10^{-39}$~J$\cdot$m$^2$, the magnon gap $\Delta /k_B = 1$~K, the magnon mean free path $l = 100$~nm, the Gilbert damping $\alpha = 10^{-4}$, and the temperature $T = 100$~K ($\lesssim $ the magnetic ordering temperature $T_c$), satisfying $\hbar \omega \ll \Delta \ll k_B T$. Based on these parameters, an estimation can be made for the thermal magnon wavelength $\lambda_T = 2 \pi \sqrt{A/k_B T} \sim 5.3$~nm, the spin diffusion constant $D \sim 1.8 \times 10^{-4}$~m$^2 \cdot$s$^{-1}$, the spin conductivity $\sigma \sim 1.4 \times 10^{36}$~J$^{-1} \cdot $s$^{-1}$, which converts to $0.04$~S in electrical units, and the spin relaxation time $\tau \sim 8.3$~ns. The characteristic wavelength is $\lambda = 4\sqrt{A/\Delta} \sim 34$~nm, the effective collision time is $\tau_0 \sim 30$~ps, and the diffusion length is $l_s = \sqrt{D \tau} \sim 1.2$~$\mu$m. These values are reasonable compared with those calculated for YIG.~\cite{Cornelissen2016}

In Fig.~\ref{fig3}, we plot the transition rate as a function of the NV resonance frequency.~\cite{Du2017,Wang2020} The diffusive description breaks down when NV distances are smaller than the magnon mean free path, or when the NV frequency is larger than the effective collision rate. Recall that we focus on the subgap noise here. For NV frequencies near or above the magnon gap, the magnetic noise will be dominated by one magnon processes.~\cite{vandersar2015}

\section{Discussion}

We briefly discuss the scaling of the NV transition rate as different power laws of the NV distance $d$ as shown in Fig.~\ref{fig1}.
First focusing on the diffusive result, 
since the form factor $q^3 e^{-2qd}$ peaks at $q \sim 1/d$, we can take the approximation 
$\chi'' (q,\omega) \sim \chi'' (1/d,\omega)$,
\(
 \Gamma (\omega,d) 
 &\sim  \frac{(\gamma \widetilde{\gamma})^2}{\beta}
\frac{\hbar^2 \chi_0}{D d^2} 
\frac{1 + (d/l_s)^2 }{[1 + (d/l_s)^2]^2+(\omega d^2/D)^2 }.
\label{eq:transition-rate-approximated}
\)
For $d \gg l_s$, the above expression scales as $d^{-4}$. Its behavior in the $l \ll d \ll l_s$ range depends on the NV frequency $\omega$, as $\Gamma \sim d^{-6}$ for $\omega \gg D/d^2$, and $\Gamma \sim d^{-2}$ for $\omega \ll D/d^2$. The zero-field NV frequency $\omega/2\pi = 2.87$~GHz is used in Fig.~\ref{fig1}, which turns out to be sufficiently large such that the $d^{-2}$ behavior is absent in the diffusive regime. We refer to Appendix. B for the plots in the low-frequency limit and at an intermediate frequency where the scaling shows a crossover from $d^{-2}$ to $d^{-6}$ as $d$ increases.
On the other hand, in the ballistic limit, as the NV approaches the material surface $d \rightarrow 0$, the magnetic noise and hence the NV relaxation rate becomes a finite constant.

Our approach is easily adaptable to magnetically ordered systems with a different dispersion relation, for example, antiferromagnetic thin films with gapped magnons. 
The results are particularly helpful for experiments on magnetic insulators exhibiting long-distance spin transport, giving a prediction for the transition rate of NV placed close to the material. 
Our theory provides a systematic framework for the NV study of magnon transport across the ballistic and diffusive regimes.
It is readily applicable for thin films of three-dimensional magnets or intrinsically two-dimensional magnetic van der Waals materials,~\cite{kim2019} while a generalization to three-dimensional materials is also straightforward.~\cite{Wang2020,weyl2021}
Future work can explore many-body effects associated with magnon-magnon and magnon-phonon interactions beyond the relaxation-time approximation, including a possibility of hydrodynamic transport aspects in ultra clean materials, which could be particularly interesting in two dimensions. A theory of coupled transport of spin and heat may be necessary at low temperatures.

\section{Conclusion}

To summarize, we have studied two-dimensional magnon transport in a magnetic thin film and derived a longitudinal dynamic spin susceptibility generally applicable across different transport regimes, which may be accessed by an NV center with a wide range of probing length scales. The resulted features in the magnetic noise at subgap frequencies are shown in the NV relaxation rate as a function of the NV distance and frequency for experimental comparison. Our theory motivates further studies of ballistic and hydrodynamic regimes in clean magnetic materials with long-range spin transport.  We have also extracted the effective diffusive parameters from the diffusive limit of the general model and discussed the dominance of low-energy magnons in two-dimensional magnon transport.

\begin{acknowledgements}
This work is supported by NSF under Grant No. DMR-1742928. H. F. acknowledges the Department of Physics and Astronomy at University of California, Los Angeles for support during the 2020 Undergraduate Summer Research Program.
We thank Chunhui Rita Du and Benedetta Flebus for helpful discussions. 
\end{acknowledgements}

\appendix
\setcounter{equation}{0}
\setcounter{figure}{0}
\renewcommand{\theequation}{A.\arabic{equation}}
\renewcommand{\thefigure}{B.\arabic{figure}}

\section{DIFFUSIVE LIMIT}
In this appendix, we show how the general expression of the dynamic spin susceptibility~(\ref{eq:full-susceptibility}) reduces to the phenomenological result~(\ref{eq:diffusive-susceptibility}) in the diffusive limit, under the assumptions of energy-dependent collision and relaxation times and a quadratic dispersion relation, as mentioned in the main text.  A ferromagnetic magnon dispersion with reduced symmetries or an antiferromagnetic magnon dispersion with a finite gap can be treated along the same line.

A few assumptions are made by taking the diffusive limit. We focus on the long-wavelength dynamics $q l \ll 1$, and further assume $q \ll k $, such that $f^0_{\vk+\vq/2} - f^0_{\vk-\vq/2} \approx ({\d f^0_\vk / \d \epsilon_{\vk}})~\vq \cdot \vv_{\vk}$, where $\mathbf{v}_\vk = \d \epsilon_\vk / \d \hbar \vk$. The following approximations can thus be taken:
\(
    \label{eq:lindhard}
    L_m(\vq, \omega)
    & \approx {\displaystyle \int} \cfrac{d^2 \vk}{ 4 \pi^2} \left(\cfrac{1}{\tau_c}\right)^{\!\!m} \cfrac{\vq \cdot \vv_{\vk}}{\vq \cdot \vv_{\vk} - \omega} \left(\cfrac{\d f^0_{\vk}}{\d \varepsilon_{\vk}}\right), \\
    N_{m,n}(\vq, \omega)
    & \approx i {\displaystyle \int} 
    \cfrac{d^2 \vk}{ 4 \pi^2} 
    \left(\cfrac{ 1 }{\tau_c}\right)^{\!\!m}
    \left(\cfrac{ 1 }{\tau_r}\right)^{\!\!n} \cfrac{1}{\vq \cdot \vv_{\vk} - \omega} \left(\cfrac{\d f^0_{\vk}}{\d \varepsilon_{\vk}}\right).
\)
To evaluate the integrals below, we will assume the low frequency limit and dominant spin-conserving scatterings $\tau_c \ll \text{min} \{ 1/\omega, \tau_r \}$, which are satisfied in the momentum window $\text{max} \{l \hbar \omega/2A,  \alpha l \Delta /A \} \ll k \ll 1/\alpha l$. For the characteristic magnons important for transport to fall in this window, the frequency needs to be constrained to $\omega \ll \pi \sqrt{A \Delta}/\hbar l$, which is precisely the frequency range $\omega \tau_0 \ll 1$ for the diffusive model to apply. For the parameters used in the main text, the upper bound of $\omega$ is roughly $35$~GHz. 
The momentum window also requires the Gilbert damping to be small: $\alpha \ll \lambda/2\pi l \sim 0.05$.
Considering the vanishing phase space factor $k dk$ in the limit of $k\rightarrow 0$, and the high-energy suppression by the Bose-Einstein distribution for $k \rightarrow \infty$, we can safely extend this momentum window to $(0, \infty)$ in evaluating the integrals.  

Expanding
\begin{align}
    \widetilde{L}_0(\vq, \omega)
    & \approx \int \frac{d^2 \vk}{ 4 \pi^2} \left[1 - i\left(\vq \cdot \vv_{\vk} \tau_c - \omega \tau_c - i \frac{\tau_c}{\tau_r}\right) \right. \nonumber\\
    & \qquad - \left.\left(\vq \cdot \vv_{\vk} \tau_c - \omega \tau_c - i \frac{\tau_c}{\tau_r}\right)^2 \right] (i\vq \cdot \vv_{\vk} \tau_c)
    \left(\frac{\partial f^0_{\vk} }{\partial \epsilon_{\vk}}\right) \nonumber \\
    & \approx \int \frac{d^2 \vk}{ 4 \pi^2} ~ (\vq \cdot \vv_{\vk} \tau_c)^2 \left(\frac{\partial f^0_{\vk} }{\partial \epsilon_{\vk}}\right) \nonumber\\
    & = \frac{1}{ 4 \pi^2}\int kdkd\phi ~ q^2 l^2 \cos^2\phi \left(\frac{\partial f^0_{\vk} }{\partial \epsilon_{\vk}}\right) \nonumber\\
    & = - \frac{q^2}{2} I_1, \label{eq:Lp0}
\end{align}
the only terms that survive $\int d^2 \vk$ are those even in $\vk$. 
We have defined the following integral constants
\(
    I_i \equiv - \frac{1}{2 \pi}\int_0^{\infty} dK~K^i \left(\frac{\partial f^0_{\vk} }{\partial \epsilon_{\vk}}\right), \label{eq:int_const_i}
\)
for notational convenience, where $K =  k l $ and $i = 1, 2, 3$.
Similarly,
\begin{align}
   \widetilde{L}_1(\vq, \omega)
    & \! \approx \! \int \frac{d^2 \vk}{ 4 \pi^2} ~ \left(\frac{1}{\tau_c}\right) (\vq \cdot \vv_{\vk} \tau_c)^2 \left(\frac{\partial f^0_{\vk} }{\partial \epsilon_{\vk}}\right)
    = - \frac{Aq^2}{\hbar l^2} I_2, \hspace{0.5cm}\label{eq:Lp1} \\
    \widetilde{N}_{1,0}(\vq, \omega)
    & \! \approx \! -\!\! \int \frac{d^2 \vk}{ 4 \pi^2} \left(\frac{\partial f^0_{\vk} }{\partial \epsilon_{\vk}}\right)
    = \frac{1}{l^2} I_1, 
 \label{eq:Np10} \\
\nonumber\\
    \widetilde{N}_{2,0}(\vq, \omega)
    & \!\approx\! -\! \! \int \! \frac{d^2 \vk}{ 4 \pi^2} \! \left( \! \frac{1}{\tau_c} \! \right) \!\!
    \left[ 1 \!+\! i \omega \tau_c \!-\! \frac{\tau_c}{\tau_r} \!-\! (\vq \cdot \vv_{\vk} \tau_c)^2 \right] \!\!\!
    \left( \! \frac{\partial f^0_{\vk}}{\partial \epsilon_{\vk}} \! \right) \nonumber \\
    & \hspace{-30pt}\! =\! \frac{2 A }{\hbar l^4}\left(1 - \frac{q^2 l^2}{2}\right) I_2 
    +  \frac{i \omega}{l^2} I_1
    - \frac{2\alpha}{\hbar l^2} \left( \frac{A}{l^2} I_3  + \Delta I_1 \right),  \\
    \nonumber\\
    \widetilde{N}_{0,1}(\vq, \omega) 
    & \! \approx \! -\! \int \frac{d^2 \vk}{ 4 \pi^2} \left(\frac{\tau_c}{\tau_r}\right)
   \left(\frac{\partial f^0_{\vk} }{\partial \epsilon_{\vk}}\right) 
    \!=\! \frac{\alpha}{l^2}  \left( I_2 + \frac{\Delta l^2}{A} I_0\right), \vspace{2pt}\label{eq:Np01} 
\end{align}
\(
    \widetilde{N}_{1,1}(\vq, \omega)
    & \! \approx \!  - \! \int \frac{d^2 \vk}{ 4 \pi^2} \left(\frac{1}{\tau_r}\right)
    \left(\frac{\partial f^0_{\vk} }{\partial \epsilon_{\vk}}\right)
    \!=\! \frac{2\alpha}{\hbar l^2} \left( \frac{A}{l^2}  I_3  + \Delta I_1 \right), \vspace{2pt}\label{eq:Np11} 
\)
and
\(
    L_1(\vq, 0) 
    \! \approx \! \int \frac{d^2 \vk}{ 4 \pi^2} \frac{1}{\tau_c}
   \left(\frac{\partial f^0_{\vk} }{\partial \epsilon_{\vk}}\right)
   \! =\! - \frac{2 A}{\hbar l^4} I_2. \label{eq:L1}
\)
Substituting Eqs.~(\ref{eq:Lp0})-(\ref{eq:L1}) into Eq.~(\ref{eq:full-susceptibility}) yields
\(
    \chi_F(q, \omega)
    & \approx \\
    & \hspace{-20pt} \frac{(A  l^2)I_2 q^2 + (2\alpha  l^2) [(A/l^2)I_3+ \Delta I_1]}{-i \hbar\omega + 2\alpha [(A/l^2)(I_3/I_1)+ \Delta]+ A (I_2/I_1) q^2},
    \label{eq:full-susceptibility_d}
\)
where leading order terms of second order in $q$ and first order in $\omega$ and $\alpha$ are kept.
Comparing Eq.~(\ref{eq:full-susceptibility_d}) with the diffusive form~(\ref{eq:diffusive-susceptibility}), the corresponding transport parameters can be extracted:
$D = (A/\hbar) (I_2/I_1)$,
$\sigma  = (A / \hbar l^2) I_2$, and
$\tau = (\hbar/2 \alpha) /[ (A/l^2) (I_3/I_1) + \Delta]$.
For $\beta \Delta \ll 1$, we get
\(
    \label{eq:int_consts}
    I_1
    & = \frac{l^2}{4 \pi A } \frac{1}{e^{\beta \Delta}-1} 
    \approx
     \frac{ l^2 }{ 4\pi A\beta\Delta}, \\
    I_2
    & = \frac{l^3}{8 \pi A} \sqrt{\frac{\pi }{\beta A}}\text{Li}_{1/2}\left(e^{-\beta\Delta}\right)
    \approx 
     \frac{l^3}{8 \beta \sqrt{A^3\Delta}}, \\
    I_3
    & \approx - \cfrac{l^4}{4\pi \beta A^2}\ln\left(\beta{\Delta }\right), 
\)
\\
where $\text{Li}_{s} (z)$ is the polylogarithm function of order $s$. These \\ yield
    $D \!\approx\! {\pi l \sqrt{A\Delta} / 2 \hbar}$, 
    $\sigma  \!\approx\! { l / 8 \hbar \beta \sqrt{A\Delta}}$, and
    $\tau  \approx -{\hbar / 2\alpha \Delta \ln\left(\beta \Delta\right)}$. \\
    
\section{ADDITIONAL PLOTS}

\begin{figure}[!ht]
\label{fig:appendix}
\includegraphics[width = \linewidth]{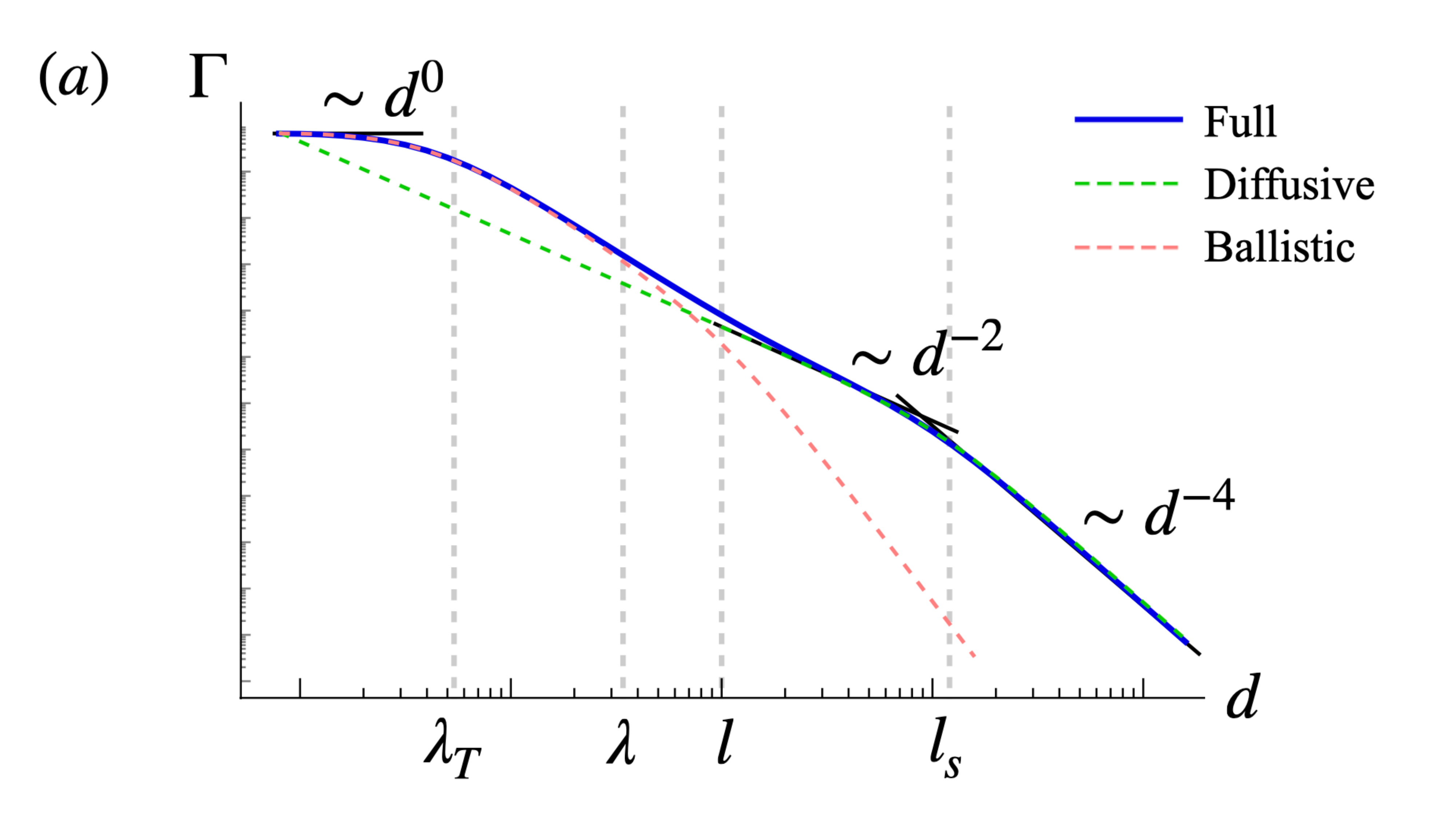} \\
\includegraphics[width = \linewidth]{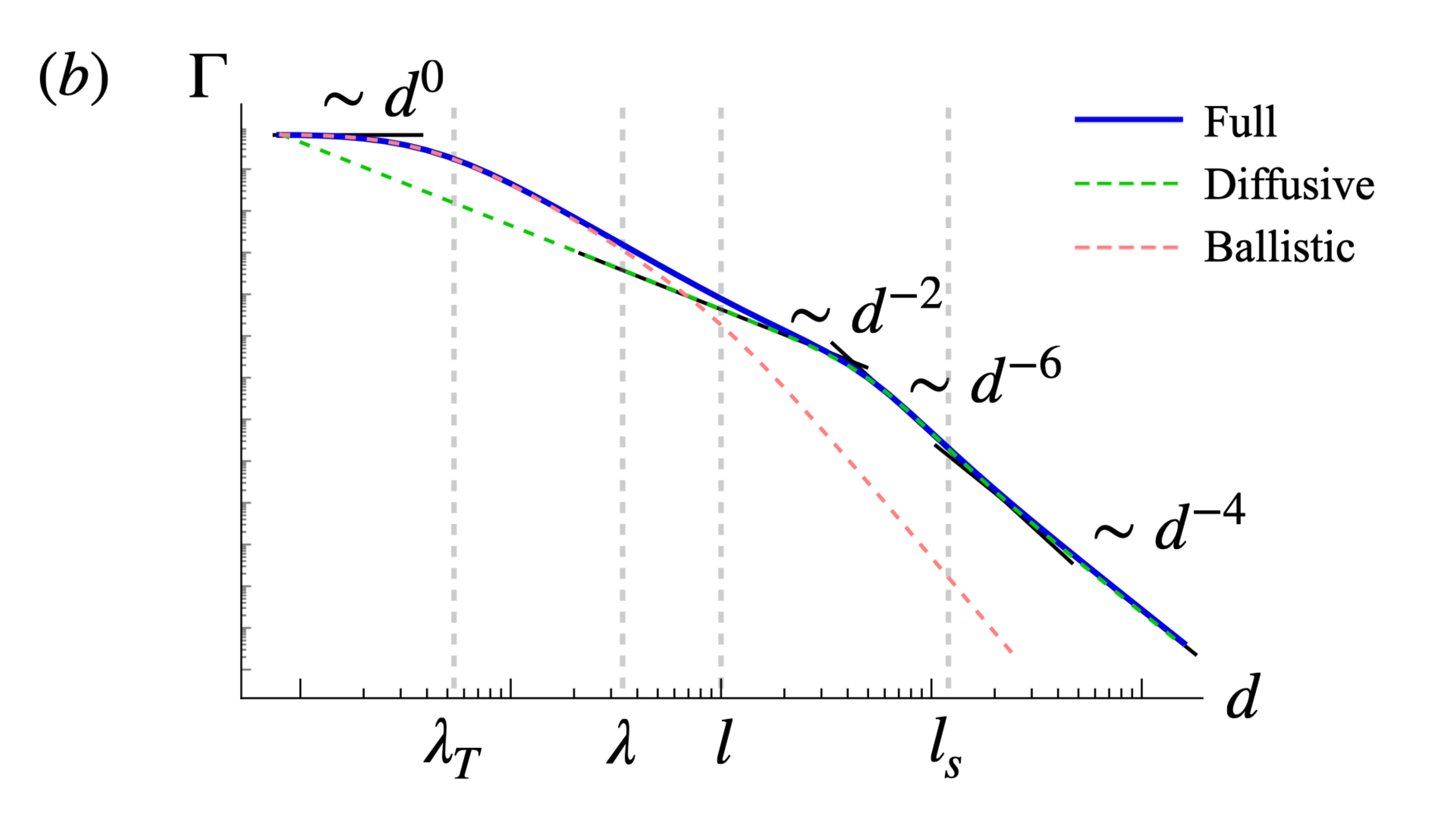} 
\caption{Log-log plot of the NV transition rate scaling with the NV distanced $d$ across  different  magnon  transport  regimes, with the NV resonance frequency $\omega$ set to be $\omega/2\pi = 0.01$~GHz (a) and $0.1$~GHz (b).}
\end{figure}

In this section, we provide additional plots of the NV relaxation rate as a function of NV distance at different choices of resonance frequency $\omega$, which tunes the relative crossover point between different power laws. As can be seen from Eq.~(\ref{eq:transition-rate-approximated}), in the limit $\omega \rightarrow 0$, the scaling of the relaxation rate $\Gamma$ with NV distance $d$ changes from $\Gamma \sim d^{-2}$ for $l \ll d \ll l_s$ to $\Gamma \sim d^{-4}$ for $d \gg l_s$, which is plotted in Fig.~\ref{fig:appendix}.1(a). On the other hand, Fig.~\ref{fig1} in the main text shows a scaling of $\Gamma \sim d^{-6}$ for $l \ll d \ll l_s$ due to the choice of a large frequency within the diffusive regime ($D/l l_s \lesssim \omega \ll 1/\tau_0$). In Fig.~\ref{fig:appendix}.1(b), an intermediate frequency value ($D/ l_s^2 < \omega < D/l l_s$) is chosen to show the scenario that exhibits all three available power laws: $\Gamma \sim d^{-2}$ for $l \ll d \ll D / \omega l_s$, $\Gamma \sim d^{-6}$ for $D / \omega l_s \ll d \ll l_s $, and $\Gamma \sim d^{-4}$ for $d \gg l_s$. 

\vspace{10pt}

\break

\bibliography{main}

\end{document}